\pdfoutput=1

\documentclass[12pt,a4paper]{article}

\usepackage{ifthen}
\newboolean{pdflatex}
\setboolean{pdflatex}{true}

\newboolean{articletitles}
\setboolean{articletitles}{true}

\newboolean{uprightparticles}
\setboolean{uprightparticles}{false}

\newboolean{inbibliography}
\setboolean{inbibliography}{false}

\def\paperauthors{LHCb collaboration}
\def\paperasciititle{Observation of the decay Bs -> chictwo K K in the
phi mass region}
\def\papertitle{ Observation of the decay $\Bsb \rightarrow \chictwo
  K^+ K^- $ in the $\Pphi$ mass region}
\def\paperkeywords{{High Energy Physics}, {LHCb}}
\def\papercopyright{\the\year\ CERN for the benefit of the LHCb collaboration}
\def\paperlicence{CC-BY-4.0 licence}
\def\paperlicenceurl{https://creativecommons.org/licenses/by/4.0/}

\usepackage[top=1in, bottom=1.25in, left=1in, right=1in]{geometry}

\usepackage{microtype}
\usepackage{xspace}
\usepackage{caption}

\usepackage{graphicx}
\usepackage{color}
\usepackage{colortbl}

\usepackage{amsmath}
\usepackage{amssymb}
\usepackage{amsfonts}
\usepackage{upgreek}

\usepackage{hyperxmp}

\usepackage[pdftex,
            pdfauthor={\paperauthors},
            pdftitle={\paperasciititle},
            pdfkeywords={\paperkeywords},
            pdfcopyright={Copyright (C) \papercopyright},
            pdflicenseurl={\paperlicenceurl}]{hyperref}

\usepackage[all]{hypcap}

\usepackage{xspace}
\usepackage{upgreek}

\def\lhcb {\mbox{LHCb}\xspace}

\def\MagUp {\mbox{\em Mag\kern -0.05em Up}\xspace}

\ifthenelse{\boolean{uprightparticles}}
{

 \def\Pphi        {\ensuremath{\upphi}\xspace}
 
 \def\Pchi        {\ensuremath{\upchi}\xspace}
 \def\Ppsi        {\ensuremath{\uppsi}\xspace}

 \def\PDelta      {\ensuremath{\Delta}\xspace}
 \def\PXi      {\ensuremath{\Xi}\xspace}
 \def\PLambda      {\ensuremath{\Lambda}\xspace}
 \def\PSigma      {\ensuremath{\Sigma}\xspace}
 \def\POmega      {\ensuremath{\Omega}\xspace}
 \def\PUpsilon      {\ensuremath{\Upsilon}\xspace}

 \def\PB      {\ensuremath{\mathrm{B}}\xspace}
 
 \def\PD      {\ensuremath{\mathrm{D}}\xspace}

 \def\PJ      {\ensuremath{\mathrm{J}}\xspace}
 \def\PK      {\ensuremath{\mathrm{K}}\xspace}

 \def\Pb      {\ensuremath{\mathrm{b}}\xspace}
 \def\Pc      {\ensuremath{\mathrm{c}}\xspace}

 \def\Pi      {\ensuremath{\mathrm{i}}\xspace}

 \def\Pp      {\ensuremath{\mathrm{p}}\xspace}

 \def\Ps      {\ensuremath{\mathrm{s}}\xspace}

}
{

 \def\Pphi        {\ensuremath{\phi}\xspace}
 
 \def\Pchi        {\ensuremath{\chi}\xspace}
 \def\Ppsi        {\ensuremath{\psi}\xspace}
 
 \mathchardef\PDelta="7101
 \mathchardef\PXi="7104
 \mathchardef\PLambda="7103
 \mathchardef\PSigma="7106
 \mathchardef\POmega="710A
 \mathchardef\PUpsilon="7107
 
 \def\PB      {\ensuremath{B}\xspace}
 
 \def\PD      {\ensuremath{D}\xspace}

 \def\PJ      {\ensuremath{J}\xspace}
 \def\PK      {\ensuremath{K}\xspace}

 \def\Pb      {\ensuremath{b}\xspace}
 \def\Pc      {\ensuremath{c}\xspace}

 \def\Pi      {\ensuremath{i}\xspace}

 \def\Pp      {\ensuremath{p}\xspace}

 \def\Ps      {\ensuremath{s}\xspace}

}

\makeatletter
\ifcase \@ptsize \relax
  \newcommand{\miniscule}{\@setfontsize\miniscule{4}{5}}
\or
  \newcommand{\miniscule}{\@setfontsize\miniscule{5}{6}}
\or
  \newcommand{\miniscule}{\@setfontsize\miniscule{5}{6}}
\fi
\makeatother

\DeclareRobustCommand{\optbar}[1]{\shortstack{{\miniscule (\rule[.5ex]{1.25em}{.18mm})}
  \\ [-.7ex] $#1$}}

\def\squark    {{\ensuremath{\Ps}}\xspace}

\def\cquark    {{\ensuremath{\Pc}}\xspace}

\def\bquark    {{\ensuremath{\Pb}}\xspace}

\def\kaon    {{\ensuremath{\PK}}\xspace}
  \def\Kbar    {{\kern 0.2em\overline{\kern -0.2em \PK}{}}\xspace}

\def\KorKbar    {\kern 0.18em\optbar{\kern -0.18em K}{}\xspace}

\def\Km      {{\ensuremath{\kaon^-}}\xspace}

\def\KS      {{\ensuremath{\kaon^0_{\mathrm{ \scriptscriptstyle S}}}}\xspace}

\def\Kstarz  {{\ensuremath{\kaon^{*0}}}\xspace}

\def\Kstar   {{\ensuremath{\kaon^*}}\xspace}

\newcommand{\phiz}{\ensuremath{\Pphi}\xspace}

  \def\Dbar    {{\kern 0.2em\overline{\kern -0.2em \PD}{}}\xspace}
\def\D       {{\ensuremath{\PD}}\xspace}

\def\DorDbar    {\kern 0.18em\optbar{\kern -0.18em D}{}\xspace}

\def\B       {{\ensuremath{\PB}}\xspace}
\def\Bbar    {{\ensuremath{\kern 0.18em\overline{\kern -0.18em \PB}{}}}\xspace}

\def\BorBbar    {\kern 0.18em\optbar{\kern -0.18em B}{}\xspace}
\def\Bz      {{\ensuremath{\B^0}}\xspace}

\def\Bu      {{\ensuremath{\B^+}}\xspace}

\def\Bd      {{\ensuremath{\B^0}}\xspace}
\def\Bs      {{\ensuremath{\B^0_\squark}}\xspace}
\def\Bsb     {{\ensuremath{\Bbar{}^0_\squark}}\xspace}

\def\jpsi     {{\ensuremath{{\PJ\mskip -3mu/\mskip -2mu\Ppsi\mskip 2mu}}}\xspace}
\def\psitwos  {{\ensuremath{\Ppsi{(2S)}}}\xspace}

\def\chiczero {{\ensuremath{\Pchi_{\cquark 0}}}\xspace}
\def\chicone  {{\ensuremath{\Pchi_{\cquark 1}}}\xspace}
\def\chictwo  {{\ensuremath{\Pchi_{\cquark 2}}}\xspace}
  \def\Y#1S{\ensuremath{\PUpsilon{(#1S)}}\xspace}

\def\proton      {{\ensuremath{\Pp}}\xspace}

\def\Xires       {{\ensuremath{\PXi}}\xspace}

\def\Lz          {{\ensuremath{\PLambda}}\xspace}
\def\Lbar        {{\ensuremath{\kern 0.1em\overline{\kern -0.1em\PLambda}}}\xspace}
\def\LorLbar    {\kern 0.18em\optbar{\kern -0.18em \PLambda}{}\xspace}

\def\Omegares    {{\ensuremath{\POmega}}\xspace}

\def\Lb      {{\ensuremath{\Lz^0_\bquark}}\xspace}

\def\Xibm    {{\ensuremath{\Xires^-_\bquark}}\xspace}

\def\Omegab    {{\ensuremath{\Omegares^-_\bquark}}\xspace}

\def\to                 {\ensuremath{\rightarrow}\xspace}

\def\CP                {{\ensuremath{C\!P}}\xspace}

\def\AT#1     {\ensuremath{A_{\mathrm{T}}^{#1}}\xspace}

\def\C#1      {\ensuremath{\mathcal{C}_{#1}}\xspace}
\def\Cp#1     {\ensuremath{\mathcal{C}_{#1}^{'}}\xspace}
\def\Ceff#1   {\ensuremath{\mathcal{C}_{#1}^{\mathrm{(eff)}}}\xspace}
\def\Cpeff#1  {\ensuremath{\mathcal{C}_{#1}^{'\mathrm{(eff)}}}\xspace}
\def\Ope#1    {\ensuremath{\mathcal{O}_{#1}}\xspace}
\def\Opep#1   {\ensuremath{\mathcal{O}_{#1}^{'}}\xspace}

\newcommand{\tev}{\ifthenelse{\boolean{inbibliography}}{\ensuremath{~T\kern -0.05em eV}}{\ensuremath{\mathrm{\,Te\kern -0.1em V}}}\xspace}
\newcommand{\gev}{\ensuremath{\mathrm{\,Ge\kern -0.1em V}}\xspace}
\newcommand{\mev}{\ensuremath{\mathrm{\,Me\kern -0.1em V}}\xspace}
\newcommand{\kev}{\ensuremath{\mathrm{\,ke\kern -0.1em V}}\xspace}
\newcommand{\ev}{\ensuremath{\mathrm{\,e\kern -0.1em V}}\xspace}
\newcommand{\gevc}{\ensuremath{{\mathrm{\,Ge\kern -0.1em V\!/}c}}\xspace}
\newcommand{\mevc}{\ensuremath{{\mathrm{\,Me\kern -0.1em V\!/}c}}\xspace}
\newcommand{\gevcc}{\ensuremath{{\mathrm{\,Ge\kern -0.1em V\!/}c^2}}\xspace}
\newcommand{\gevgevcccc}{\ensuremath{{\mathrm{\,Ge\kern -0.1em V^2\!/}c^4}}\xspace}
\newcommand{\mevcc}{\ensuremath{{\mathrm{\,Me\kern -0.1em V\!/}c^2}}\xspace}

\def\mum  {\ensuremath{{\,\upmu\mathrm{m}}}\xspace}

\def\invfb   {\ensuremath{\mbox{\,fb}^{-1}}\xspace}

\def\ps   {\ensuremath{{\mathrm{ \,ps}}}\xspace}

\newcommand{\stat}{\ensuremath{\mathrm{\,(stat)}}\xspace}
\newcommand{\syst}{\ensuremath{\mathrm{\,(syst)}}\xspace}

\newcommand{\chisq}{\ensuremath{\chi^2}\xspace}

\newcommand{\chisqip}{\ensuremath{\chi^2_{\text{IP}}}\xspace}

\def\gsim{{~\raise.15em\hbox{$>$}\kern-.85em
          \lower.35em\hbox{$\sim$}~}\xspace}
\def\lsim{{~\raise.15em\hbox{$<$}\kern-.85em
          \lower.35em\hbox{$\sim$}~}\xspace}

\def\sPlot{\mbox{\em sPlot}\xspace}

\def\ptot       {\mbox{$p$}\xspace}
\def\pt         {\mbox{$p_{\mathrm{ T}}$}\xspace}

\def\evtgen     {\mbox{\textsc{EvtGen}}\xspace}

\def\geant      {\mbox{\textsc{Geant4}}\xspace}

\def\photos     {\mbox{\textsc{Photos}}\xspace}

\def\pythia     {\mbox{\textsc{Pythia}}\xspace}

\def\tell1  {TELL1\xspace}
\def\ukl1   {UKL1\xspace}

\usepackage{cite}
\usepackage{mciteplus}

\begin{document}

\renewcommand{\thefootnote}{\fnsymbol{footnote}}
\setcounter{footnote}{1}

\begin{titlepage}
\pagenumbering{roman}

\vspace*{-1.5cm}
\centerline{\large EUROPEAN ORGANIZATION FOR NUCLEAR RESEARCH (CERN)}
\vspace*{1.5cm}
\noindent
\begin{tabular*}{\linewidth}{lc@{\extracolsep{\fill}}r@{\extracolsep{0pt}}}
\vspace*{-1.5cm}\mbox{\!\!\!\includegraphics[width=.14\textwidth]{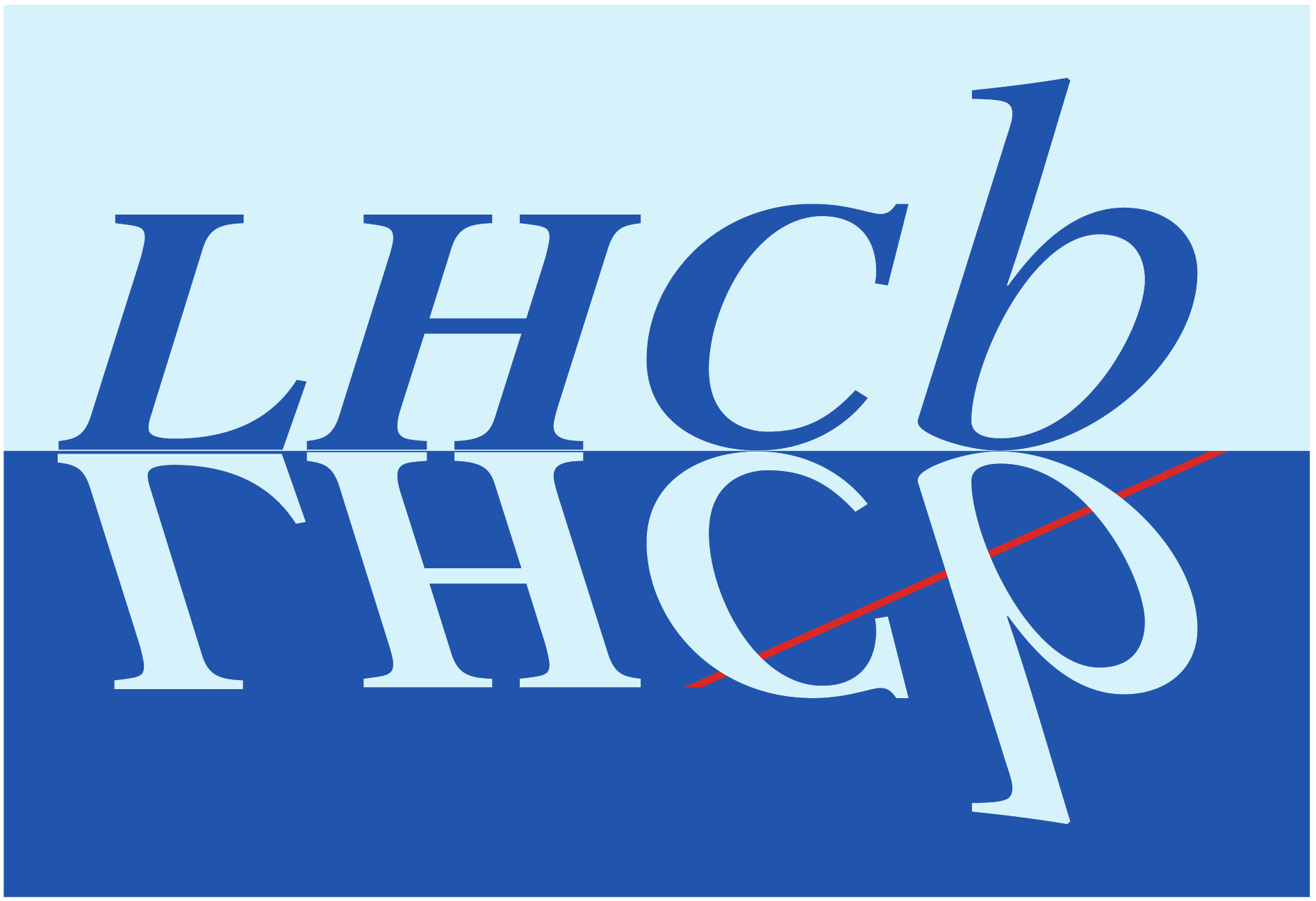}} & &
\\
 & & CERN-EP-2018-160 \\
 & & LHCb-PAPER-2018-018 \\
 & & \today \\
 & & \\
\end{tabular*}

\vspace*{4.0cm}
{\normalfont\bfseries\boldmath\huge
\begin{center}
  \papertitle
\end{center}
}

\vspace*{2.0cm}

\begin{center}
\paperauthors\footnote{Authors are listed at the end of this paper.}
\end{center}

\vspace{\fill}

\begin{abstract}
  \noindent
  The $\Bsb \rightarrow \chi_{c2} K^+ K^- $ decay mode is observed and
  its branching fraction relative to the corresponding $\chicone$ decay
 mode, in a $\pm 15 \mevcc$ window around the $\phi$ mass, is found to be
\begin{displaymath}
\frac{\mathcal{B}(\Bsb  \rightarrow \chi_{c2} K^+ K^-) }{ \mathcal{B}(\Bsb \rightarrow \chi_{c1} K^+
K^-)} =  (17.1 \pm 3.1 \pm  0.4 \pm 0.9)\%,
\end{displaymath}
where the first uncertainty is statistical, the second systematic and
the third due to the knowledge of the branching fractions of radiative
$\chi_c$ decays. The decay mode $\Bsb \rightarrow \chi_{c1} K^+ K^- $ allows the $\Bs$ mass to be measured as
\begin{displaymath}
m(\Bs) = 5366.83 \pm 0.25 \pm 0.27 \, \textrm{MeV}/c^2,
\end{displaymath}
 where the first uncertainty is statistical and the second
 systematic. A combination of this result with other LHCb
 determinations of the $\Bs$ mass is made.
\end{abstract}

\vspace*{1.0cm}

\begin{center}
 Published as JHEP 08 (2018) 191.
\end{center}

\vspace{\fill}

{\footnotesize
\centerline{\copyright~\papercopyright. \href{\paperlicenceurl}{\paperlicence}.}}
\vspace*{2mm}

\end{titlepage}

\newpage
\setcounter{page}{2}
\mbox{~}

\cleardoublepage

\renewcommand{\thefootnote}{\arabic{footnote}}
\setcounter{footnote}{0}

\pagestyle{plain}
\setcounter{page}{1}
\pagenumbering{arabic}


\section{Introduction}
\label{sec:Introduction}
Studies of two-body $\bquark$-hadron decays to final states containing a hidden
charm meson such as a $\chi_{cJ}$ state ($J = 0,1,2$) provide powerful probes of the strong interaction. These
decays proceed predominantly  via a colour-suppressed
$\bquark \rightarrow \cquark \bar{\cquark}\squark$
transition. Theoretically, such decays are often studied in the
factorization approach \cite{Diehl:2001xe,Beneke:1998ks}. It is predicted, in the absence of final-state interactions, that decays to
spin-0 and 2 charmonium states are highly suppressed compared to
decays to spin-1 states \cite{Diehl:2001xe}. Experimentally,
factorization has been observed to hold for $\Bu \rightarrow \chi_{c1,c2}
K^+$ decays,\footnote{The inclusion of charge-conjugate processes is implied
throughout this paper.} for which the Belle collaboration reported ${\mathcal{B}(\Bu \rightarrow \chictwo
K^+)/\mathcal{B}(\Bu \rightarrow \chicone K^+)  =
(2.25^{+0.73}_{-0.69}  \stat \pm 0.17 \syst) \times 10^{-2}}$~\cite{Bhardwaj:2011dj}. In other modes, less suppression is
observed. For example, the LHCb collaboration has measured
$\mathcal{B}(\Bz \rightarrow \chictwo K^{*}(892)^0)/ \mathcal{B}(\Bz
\rightarrow \chicone K^{*}(892)^0) = (17.1 \pm 5.0 \stat \pm  1.7
\syst \pm 1.1 \, (\mathcal{B}))\times 10^{-2}$ \cite{LHCb-PAPER-2013-024}, where the third uncertainty is due to
the knowledge of external branching fractions, and the Belle
collaboration has measured ${\mathcal{B}(\Bu \rightarrow \chictwo K^+ \pi^+
\pi^-)/\mathcal{B}(\Bu \rightarrow \chicone K^+ \pi^+ \pi^-) = 0.36 \pm
0.05}$~\cite{Bhardwaj:2015rju}, where the total uncertainty is quoted.
Even more strikingly, the LHCb collaboration reported \cite{LHCb-PAPER-2017-011}
${\mathcal{B}(\Lb \rightarrow  \chictwo  p
K^-)/\mathcal{B}(\Lb \rightarrow  \chicone  p K^-) = 1.02 \pm
0.10 \stat \pm 0.02 \syst \pm 0.05 \,  (\mathcal{B}).}$ These observations are
difficult to reconcile with the factorization hypothesis. It is thus interesting to probe this ratio with other exclusive decay modes.

\begin{figure}[htb!]
\begin{center}
\resizebox{4.2in}{!}{\includegraphics{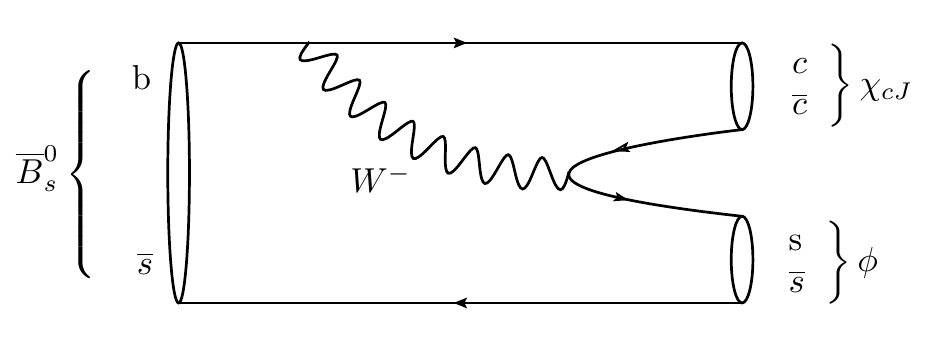}}
\caption{\small Tree-level Feynman diagram for the $\Bsb
  \rightarrow \chi_{cJ} \phi$ decay mode.}
\label{fig1}
\end{center}
\end{figure}

In this paper, the decay $\Bsb \rightarrow \chictwo K^+ K^-$ (with charge
conjugation implied) with
$\chictwo \rightarrow \jpsi \gamma$ and $\jpsi \rightarrow \mu^+ \mu^-$ is
observed using the LHCb data set collected in $pp$ collisions up to
the end of 2016. The data corresponds to an integrated luminosity of $3.0
\invfb$ collected at centre-of-mass energies of 7 and 8~TeV during 2011
and 2012, together with $1.9 \invfb$ collected at a
centre-of-mass energy of 13~TeV during 2015 and 2016. This analysis focuses
on the low $K^+K^-$ mass region, where the $\Bsb \rightarrow \chi_{cJ} K^+ K^-$
decay is expected to be dominated by the decay of an intermediate
$\phi$ meson, as shown in Fig.~\ref{fig1}. The same data set allows a
measurement of the $\Bs$
mass with high precision due to the relatively small energy release. These studies build on the previous observation of the $\Bsb \rightarrow \chicone \phi$ mode \cite{LHCb-PAPER-2013-024}.

\section{Detector and simulation}
\label{sec:Detector}
The \lhcb detector~\cite{Alves:2008zz,LHCb-DP-2014-002} is a
single-arm forward spectrometer covering the \mbox{pseudorapidity}
range $2<\eta <5$, designed for the study of particles containing
\bquark or \cquark quarks. The detector includes a high-precision
tracking system consisting of a silicon-strip vertex detector
surrounding the $pp$ interaction
region~\cite{LHCb-DP-2014-001}, a large-area silicon-strip
detector located upstream of a dipole magnet with a bending power of
about $4{\mathrm{\,Tm}}$, and three stations of silicon-strip
detectors and straw drift tubes~\cite{LHCb-DP-2013-003} placed
downstream of the magnet. The tracking system provides a measurement
of the momentum, \ptot, of charged particles with a relative uncertainty
that varies from 0.5\% at low momentum to 1.0\% at 200\gevc. The
momentum scale is calibrated using samples of $\jpsi \rightarrow \mu^+
\mu^-$ and $\Bu \rightarrow \jpsi K^+$ decays collected concurrently
with the data sample used for this analysis \cite{LHCb-PAPER-2012-048,LHCb-PAPER-2013-011}. The relative accuracy of this
procedure is estimated to be $3 \times 10^{-4}$ using samples of other
fully reconstructed $\bquark$-hadron, narrow-$\PUpsilon$, and
$\KS$ decays. The minimum distance of a track to a primary vertex (PV), the impact
parameter (IP), is measured with a resolution of $(15+29/\pt)\mum$,
where \pt is the component of the momentum transverse to the beam,
in\,\gevc.

Different types of charged
hadrons are distinguished using information from two ring-imaging Cherenkov (RICH)
detectors. Photons, electrons and hadrons are
identified by a calorimeter system consisting of scintillating-pad and
preshower detectors, an electromagnetic calorimeter and a hadronic
calorimeter. Muons are  identified by a system composed of alternating
layers of iron and multiwire proportional chambers~\cite{LHCb-DP-2012-002}.

The online event selection is performed by a
trigger~\cite{LHCb-DP-2012-004}, which consists of a hardware stage,
based on information from the calorimeter and muon
systems, followed by a software stage, where  a full event
reconstruction is made. Candidate events are required to pass the
hardware trigger, which selects muon and dimuon candidates with high
$\pt$ based upon muon-system information. The subsequent software trigger is composed of two stages. The first performs a
partial event reconstruction and requires events to have two
well-identified oppositely charged muons with an invariant mass larger
than $2.7 \gevcc$. The second stage performs a full event
reconstruction.  Events are retained for further processing if they
contain a $\jpsi \rightarrow \mu^+ \mu^-$ candidate. The distance
between the decay vertex of the $\jpsi$ and each PV, divided by its
uncertainty, is required to be larger than three.

To study the properties of the signal and the most important
backgrounds, simulated $pp$ collisions are generated using
\pythia~\cite{Sjostrand:2006za,*Sjostrand:2007gs}  with a specific
\lhcb configuration~\cite{LHCb-PROC-2010-056}.  Decays of hadronic
particles are described by \evtgen~\cite{Lange:2001uf}, in which
final-state radiation is generated using
\photos~\cite{Golonka:2005pn}. The
interaction of the generated particles with the detector, and its
response, are implemented using the \geant
toolkit~\cite{Allison:2006ve, *Agostinelli:2002hh} as described in
Ref.~\cite{LHCb-PROC-2011-006}. Other sources of background, such as
those from $\bquark \rightarrow \psitwos$ transitions, where the $\psitwos$
decays radiatively to a $\chi_{cJ}$ meson, are studied using the {\mbox{\textsc{RapidSim}}\xspace} fast simulation package \cite{Cowan:2016tnm}.

\section{Selection}
\label{sec:selection}
A two-step procedure is used to optimize the selection of $\Bsb \rightarrow
\chi_{c1,c2} K^+ K^-$ candidates. These studies use simulation samples together with the
high-mass sideband of the data, $5550 < m(\chictwo K^+ K^-) < 6150
\mevcc$, which is not used for subsequent analysis. In a first step, loose selection criteria are applied to
reduce the background significantly whilst retaining
high signal efficiency. Subsequently, a
multivariate selection is used to reduce further the combinatorial
background.

The selection starts from a pair of oppositely charged particles,
identified as muons, that form a common decay vertex. Combinatorial
background is suppressed by requiring that the $\chisqip$ of the muon candidates, defined as the difference between
the $\chisq$ of the PV reconstructed with and without the considered
particle, be larger than four for all
reconstructed PVs.  The
  invariant mass of the dimuon candidate must be within $50 \mevcc$ of the known $\jpsi$ mass
\cite{PDG2016}.

Photons are selected from well-identified neutral clusters, reconstructed in the electromagnetic
calorimeter~\cite{LHCb-DP-2014-002}, that have a transverse energy in
excess of $700 \mevc$. Selected  $\jpsi$ and photon candidates are
combined to form $\chi_{c1,c2}$ candidates. The invariant mass of the combination, obtained from a kinematic fit
\cite{Hulsbergen:2005pu}  with a $\jpsi$ mass constraint \cite{PDG2016},  is required to be
within the range $3400$--$3700 \mevcc$.

Pairs of oppositely charged kaons with $\pt > 200 \mevc$ and displaced from all PVs ($\chisqip > 4$) are selected. Good kaon
identification is achieved by using information from the RICH
detectors. This is combined with kinematic and track quality information using neural networks which provide a response that varies
between 0 and 1 for each of the different mass hypotheses: kaon ($\mathcal{P}^K$),
pion ($\mathcal{P}^{\pi}$), and proton ($\mathcal{P}^{\proton}$). The
closer to one this value is, the higher the likelihood that the particular mass hypothesis is
correct. The chosen requirements on these variables have an
efficiency of $(86.8 \pm 0.2)\%$  and ($86.4 \pm 0.2)\%$ for the
$\Bsb \rightarrow \chicone K^+ K^-$ and $\Bsb \rightarrow \chictwo K^+
K^-$ modes, respectively, where the uncertainty is due the size of the
available simulation samples. The invariant mass of the selected kaon pair is required to be within $ 15
\mevcc$ of the known value of the $\phi$ mass \cite{PDG2016}. These criteria substantially reduce
background from $\Kstar(892)^0$ decays where a pion is misidentified as a
kaon. To reduce background from $\Lb$ decays to excited $\Lz$ states a loose proton veto is
applied to both kaon candidates.

The $\chi_{c1,c2}$ candidate is combined with the pair of kaons to make
a candidate $\Bsb$ meson, which is associated to the PV giving the minimum \chisqip.  A kinematic fit is performed in which the
candidate is constrained to point to this PV and the dimuon mass is
constrained to the known value of the $\jpsi$ mass \cite{PDG2016}. The
reduced $\chi^2$~of this fit is required to be less than five. Combinatorial background is further reduced by requiring the
decay time of the $\Bsb$ candidate to be larger than $0.3 \ps$ and
its $\chisqip$ to be less than 20.

Several vetoes are applied to remove background from fully
reconstructed $b$-hadron decay modes.  By
combining kinematic and particle-identification information it is possible
to impose requirements that are almost fully efficient for signal
decays. The upper-mass sideband is found to be polluted by fully reconstructed
$\bquark$-hadron decays where a random photon is added. The most
important of these is the $\Bsb \rightarrow \jpsi
\phi$ decay mode. This is removed by
rejecting candidates in which the  reconstructed $\jpsi K^+ K^-$
invariant mass, calculated with a $\jpsi$ mass constraint, is
within $18 \mevcc$ ($\pm 3 \sigma$) of the known $\Bs$ mass \cite{PDG2016}. A similar
background is possible from the $\Bz \rightarrow \jpsi K^+
\pi^-$ decay mode where the pion is misidentified as a kaon. The
candidate is rejected if either of the two
possible $\jpsi K^+ \pi^-$ masses is within $18 \mevcc$ of the known
$\Bd$ mass. These two requirements reject a
negligible number of signal decays. Finally, candidates in which
either of the kaons is consistent with being a proton
($\mathcal{P}^{\proton} > \mathcal{P}^{K}  $) are rejected if the reconstructed $\jpsi \proton K^-$ mass is
within $18 \mevcc$ of the  known $\Lb$ mass. The efficiency of
this veto is $99.3\%$ for signal decays. Background from the $\Lb \rightarrow \chi_{c1,c2} \proton K^-$
decay mode peaks in the signal regions. Therefore, a veto is
applied to each kaon candidate in turn. The candidate is rejected if the $\chi_{c1, c2}
\proton K^-$ mass is within $10 \mevcc$ of the $\Lb$ mass (a
$\pm 2 \sigma$ window) and the proton well identified.
\begin{figure}[t!]
\begin{center}
\includegraphics[width=0.495\textwidth]{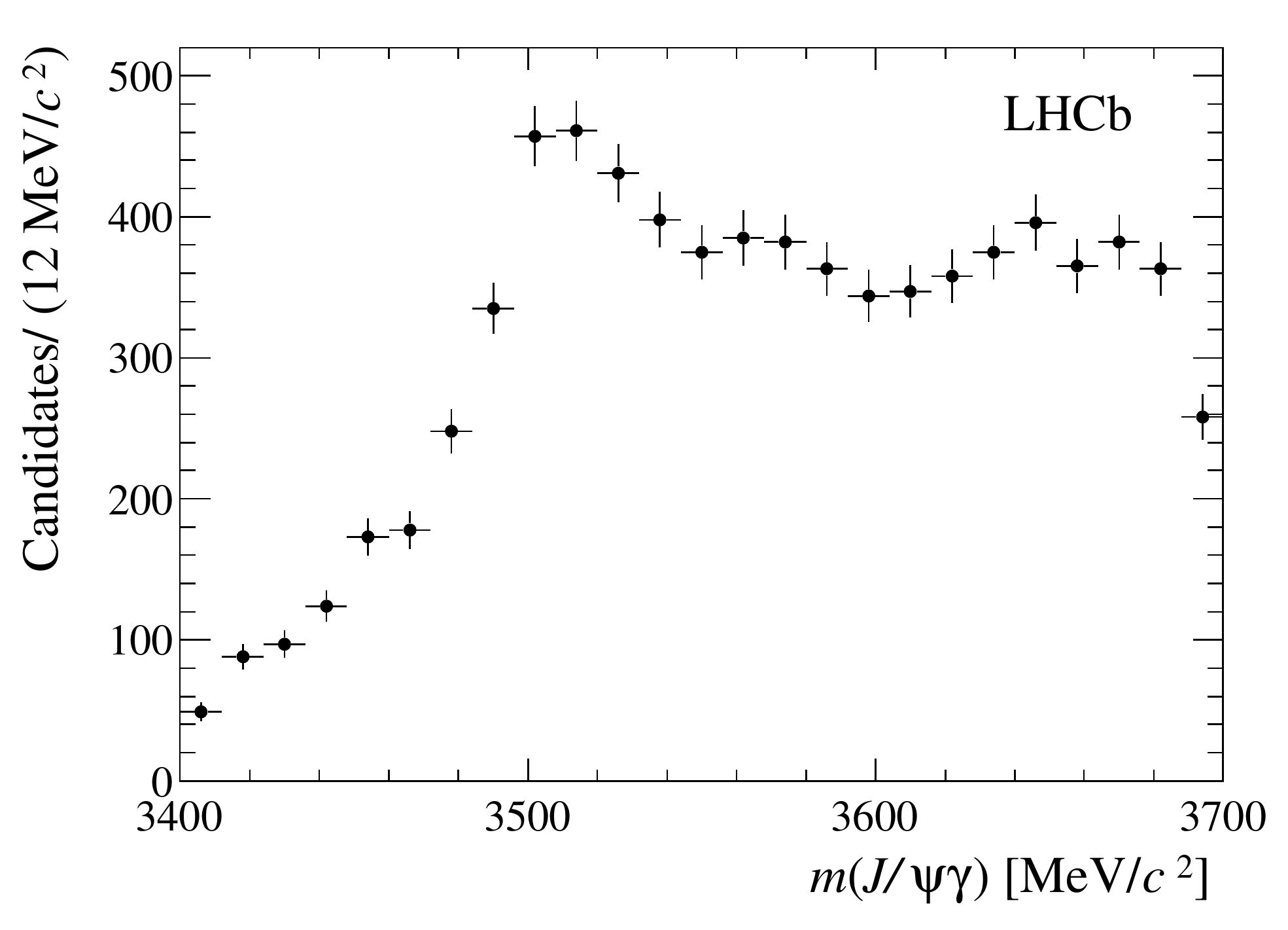}
\includegraphics[width=0.495\textwidth]{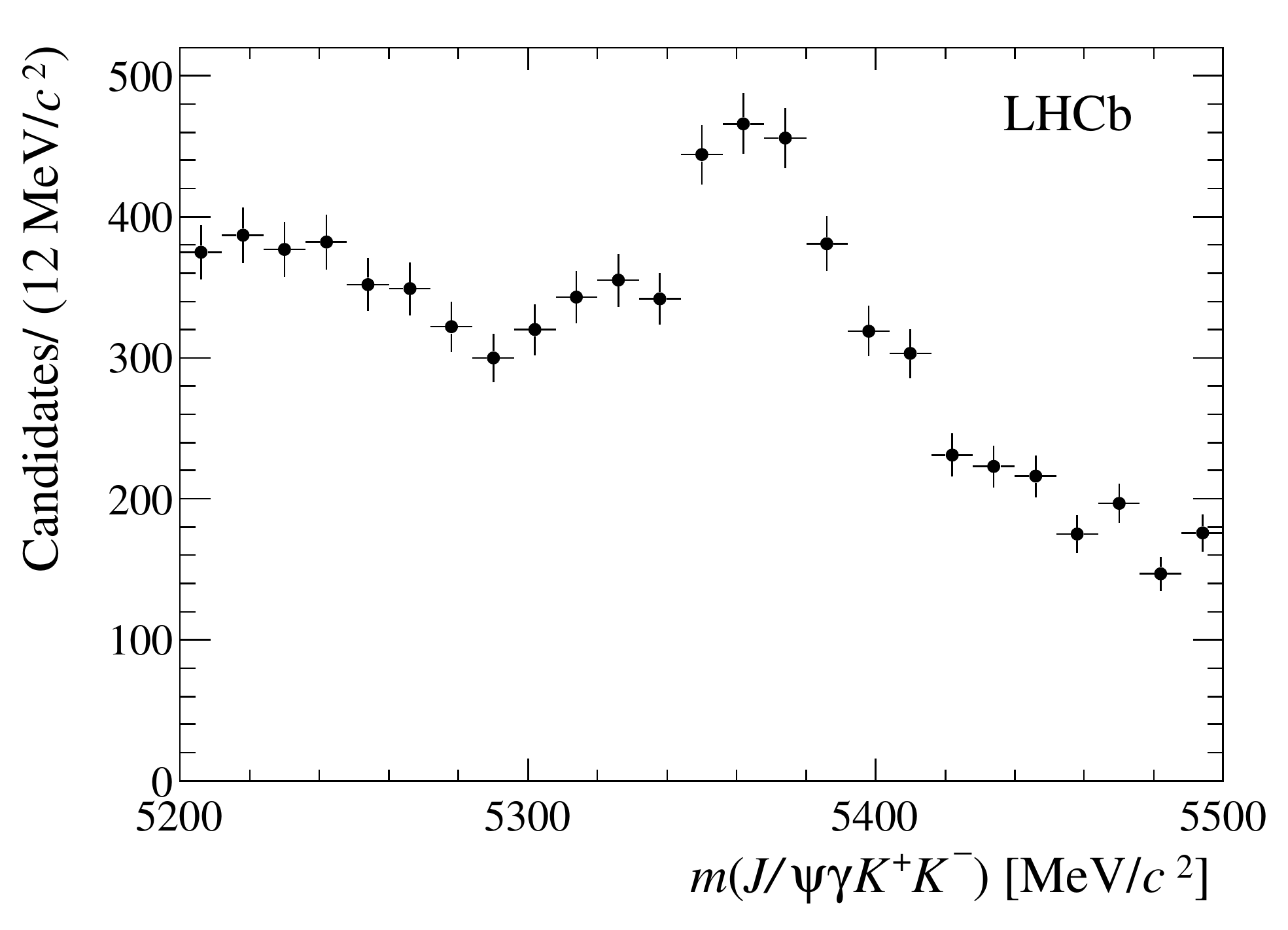}
\caption{\small Invariant-mass distributions of (left) $\jpsi  \gamma$
  and (right) $\jpsi \gamma
  K^+ K^-$  after the loose selection criteria. }
\label{fig2}
\end{center}
\end{figure}
After these requirements a broad signal is seen in the $\chi_{c1,c2}$ mass region and the $\Bsb
\rightarrow \jpsi \gamma K^+ K^-$
decay mode is observed (Fig.~\ref{fig2}) above a large combinatorial
background.

\begin{figure}[thb!]
\begin{center}
\resizebox{3.1in}{!}{\includegraphics{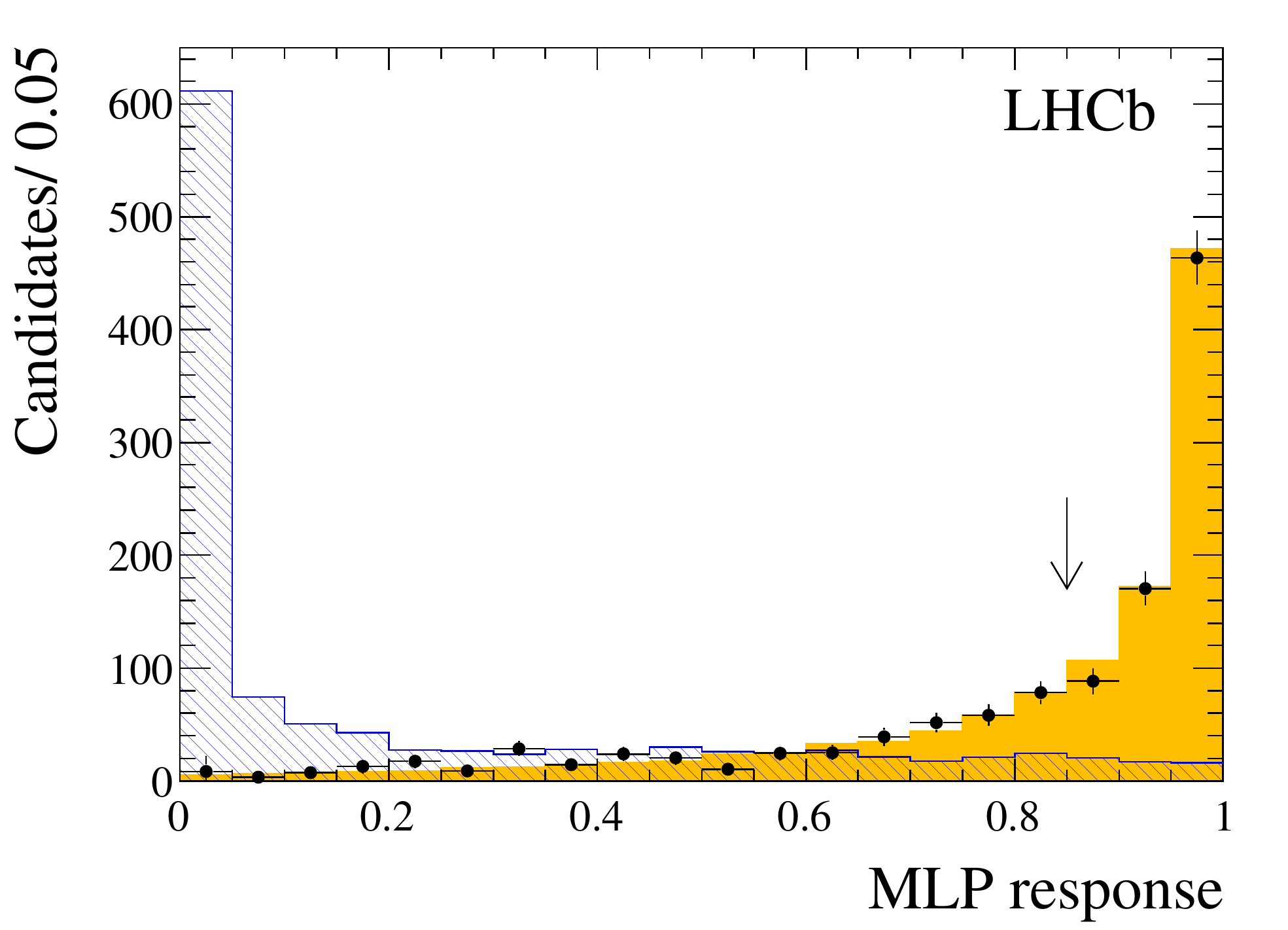}}
\caption{\small MLP response for  (solid yellow) the $\Bsb \rightarrow \chicone K^+
  K^-$ simulation sample, (hashed blue) high-mass
  sideband and (black points) the background subtracted $\Bsb \rightarrow \chicone K^+ K^-$ signal in data. The histogram areas are normalized to the number of $\Bsb
  \rightarrow \chicone K^+ K^-$ candidates observed in data after the
   loose selection. The arrow indicates the selected threshold. }
\label{fig3}
\end{center}
\end{figure}

The second step of the selection process is based
on a multilayer perceptron (MLP) classifier \cite{Hocker:2007ht,*TMVA4},  trained
using the $\Bsb \rightarrow \chicone K^+
  K^-$  and $\Bsb \rightarrow \chictwo K^+
  K^-$ simulated signal samples and the high-mass sideband of the
data. As input, the classifier uses ten variables,
related to the displacement of the candidate from the associated PV  and kinematics,
that show good agreement between data and simulation. Figure
\ref{fig3} shows the output of the MLP for the training samples
and the $\Bsb \rightarrow \chicone K^+ K^-$ signal in data where
the background is subtracted using the \sPlot technique
\cite{Pivk:2004ty}. The MLP gives excellent separation between
signal and background and shows good agreement between data and
simulation.

The requirement on the MLP output is chosen to maximize the
figure of merit $\epsilon/(a/2 + \sqrt{N_B})$ \cite{Punzi:2003bu}, where $\epsilon$ is the
signal efficiency for the $\chictwo$ mode obtained from the simulation, $a = 5$ is the
target signal significance, and $N_B$ is the background yield in a
$\pm 25 \mevcc$ window centred on the known $\Bs$ mass \cite{PDG2016} estimated from the sideband. The chosen threshold of
0.85 has an efficiency of $(65.1 \pm 0.3) \%$ for the $\Bsb\rightarrow \chicone K^+ K^-$ decay mode and $(66.1
\pm 0.3) \%$ for the $\Bsb\rightarrow \chictwo K^+ K^-$ decay mode  whilst rejecting ($96.0 \pm 0.3) \%$ of the combinatorial background.

\section{Mass fit}
\label{sec:massfit}
\begin{figure}[tb!]
\begin{center}
\includegraphics[width=0.495\textwidth]{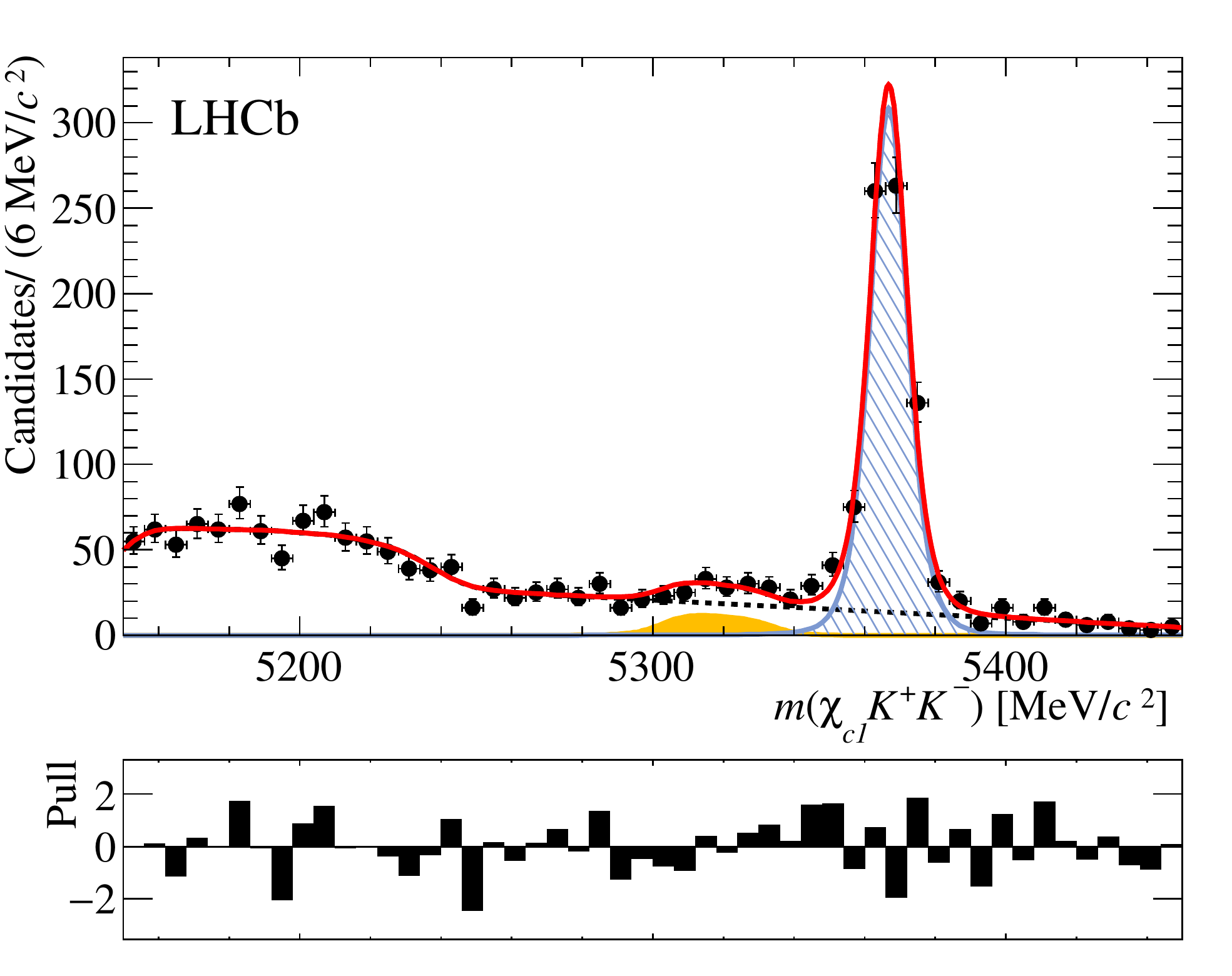}
\includegraphics[width=0.495\textwidth]{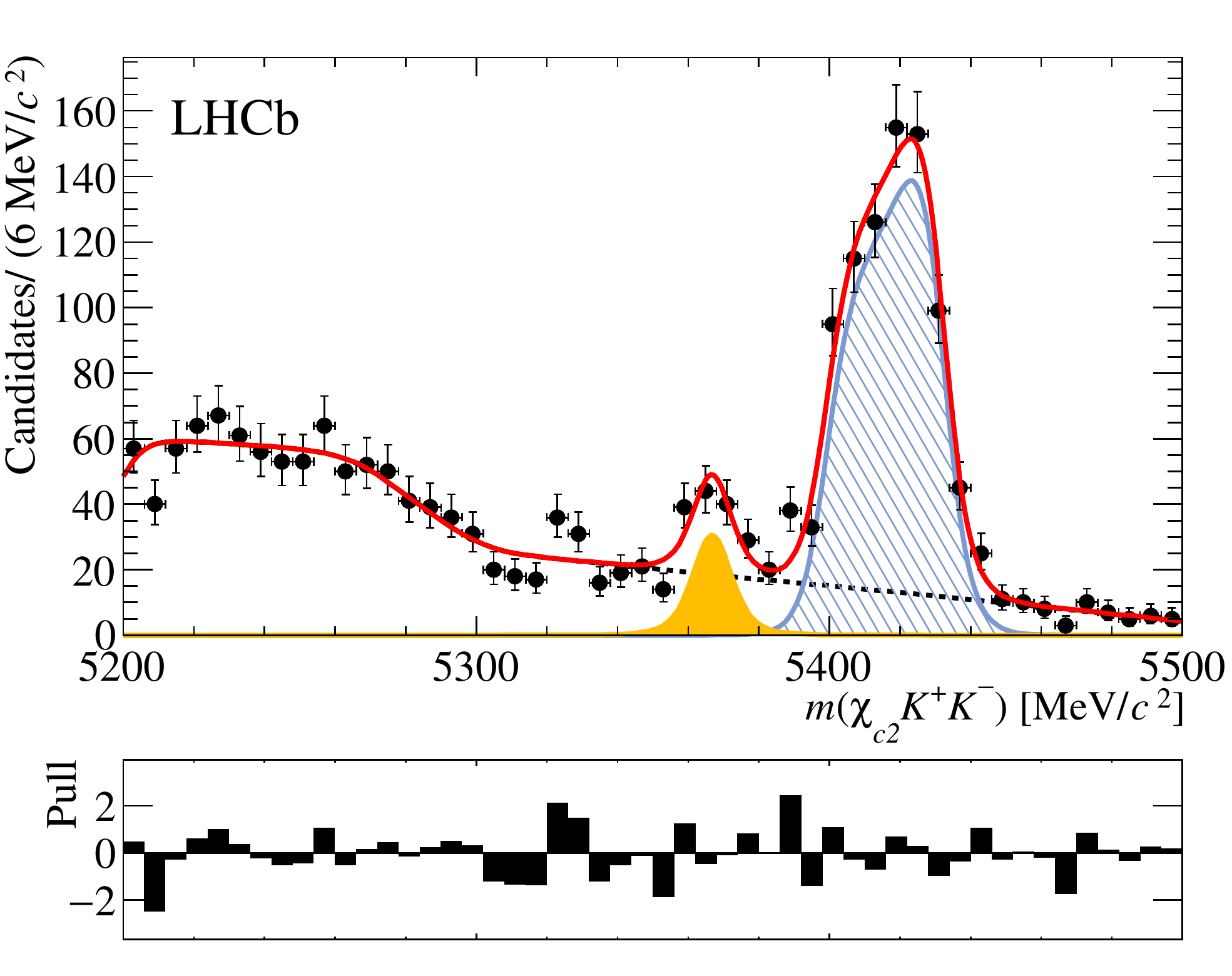}
\caption{ Invariant-mass distributions of selected candidates for
  (left) the $\chicone
  K^+ K^-$ fit and (right) the $\chictwo K^+ K^-$ fit. The total fitted function is
  superimposed (solid red line) together with the (blue hashed area)
  $\chicone$ component,  (solid yellow) $\chictwo$ component and (dashed black line) the
  background component.  The pull,
  \textit{i.e.} the difference between the observed and fitted value divided by the uncertainty,
  is shown below each of the plots.}
\label{fig4}
\end{center}
\end{figure}

The energy resolution of the LHCb calorimeter results in an
invariant-mass resolution for the  $\chicone$ and $\chictwo$  states of about
$50 \mevcc$. This makes it difficult to separate the two states
based on the $\jpsi \gamma$ invariant mass alone. To improve the mass
resolution, the approach used in previous LHCb analyses
\cite{LHCb-PAPER-2013-024,LHCb-PAPER-2017-011} is followed.  Two
kinematic fits are made to the dataset in which constraints
are applied to ensure the pointing of the candidate to the associated
primary vertex, on the $\jpsi$ mass and either on the $\chictwo$ or $\chicone$
mass. Owing to the small radiative branching fraction any contribution
from the $\Bsb \to \chiczero K^+ K^-$ decay mode can be ignored.  As can be seen in Fig.~\ref{fig4} the two components are then separable from the $\Bs$ invariant
mass calculated from this fit. A mass model for the $\Bsb \rightarrow \chi_{c1,c2} K^+ K^-$
signal is developed using the simulation.  This factorizes the
observed width of the mass distribution into a component related to
the constraints and a component related to the detector resolution.

The effect of applying the $\chictwo$ mass constraint can be seen as
follows.\footnote{The same formalism applies for a $\chicone$ mass constraint.} To
satisfy the constraint, the kinematic fit adjusts the photon momentum,
which is the most poorly measured quantity,
by a factor, $1-\alpha$, where
\begin{equation*}
\alpha = \frac{m_{\chictwo}^2  - m_{\jpsi \gamma}^2 }{ m_{\jpsi}^2 -
  m_{\jpsi \gamma}^2}
\end{equation*}
and $m_{\chictwo}$ and $m_{\jpsi}$ are the known values of the
$\chictwo$ and $\jpsi$ masses \cite{PDG2016}, respectively.
For each event in the simulation the value of $\alpha$ can be
calculated using the generated four-momenta. Then the generated four-momentum of
the photon is scaled by $1-\alpha$ and the four-momentum of the
$\Bsb$ meson recalculated. In this way the effect of the constraint is
emulated. For genuine $\Bsb \rightarrow
\chictwo K^+ K^-$ decays, applying a $\chictwo$ mass constraint transforms
the true $\Bs$ invariant-mass distribution from a $\delta$-function to a
Breit-Wigner distribution whose width is equal to the natural width of the $\chictwo$
state. In the case of  genuine $\Bsb \rightarrow \chicone K^+ K^-$ decays the distribution is shifted upwards in mass by
an amount equal to the mass splitting between the $\chictwo$ and
$\chicone$ states and is broadened. The RMS of the resulting distribution is $9.5 \mevcc$,
which allows the separation of the $\chicone$ and $\chictwo$
components.

To obtain the mass models for the $\chicone$ and $\chictwo$
components, the distributions described above are convolved with a
resolution function that accounts for the uncertainty in the measurement of the
kaon four-momenta by the tracking system. Using the simulation, the
resolution model is found to be well
described by a Student's t-distribution which has two resolution
parameters: $s$ which describes the core and $n$ which controls the
tail of the distribution. As part of the systematic studies, the
following alternative resolution models are also considered: Gaussian,
sum of two Gaussians, double-sided Crystal Ball
\cite{Skwarnicki:1986xj,LHCB-Paper-2011-013} and Bukin \cite{Bukin} functions. The advantage of factorizing the
mass distribution in this way is that it leads to a model where all parameters
can be fixed from physics considerations apart from an overall resolution scale factor, $s_f$, that accounts
for differences between data and simulation. The simulation is tuned
to match the mass resolution seen in data for the $\Bu \rightarrow
\jpsi K^+$, $\Bz \rightarrow \jpsi K^+ \pi^-$ and $\Bsb \rightarrow
\jpsi \phi $ decay modes with a precision of $5
\%$.  The validity of this tuning  for $\Bsb \rightarrow \chi_{c1, c2}
K^+ K^-$ decays is cross-checked using $\Lb \rightarrow
\chi_{c1,2} p K^-$ candidates, which have a similar topology, selected using
the criteria described in Ref.\cite{LHCb-PAPER-2017-011}. Similar
agreement between data and simulation is found and consequently in this analysis a Gaussian constraint is
applied, $s_f = 1.00 \pm 0.05$.

After the selection described in Sec. \ref{sec:selection} three
sources of background remain and are included in the
mass fit. By default, combinatorial background is modelled by a
first-order polynomial. Both a power law and an exponential function are
considered as systematic variations. Partially reconstructed background from $\Bsb \rightarrow
\psitwos K^+ K^-$ decays, with the subsequent decay $\psitwos \rightarrow
\chi_{cJ} \gamma$, is studied using  {\mbox{\textsc{RapidSim}}\xspace} and the
resulting template is added to the fit. The residual background from $\Bz \rightarrow
\chi_{c1,c2} \Kstar(892)^0$ and partially reconstructed $\Bz \rightarrow
\psitwos \Kstar(892)^0$ decays is estimated to be $7 \pm 2$ candidates and
is included as a fixed component in the fit  with the shape modelled
using the simulation.

Extended unbinned maximum likelihood fits are applied separately to
the invariant-mass distribution of selected candidates with either a $\chicone$ or $\chictwo$ mass constraint
applied. The former fit (refered to as the $\chicone K^+ K^-$ fit) is used to make further
cross-checks of the mass resolution and to determine the $\Bs$
mass. The latter (refered to as the $\chictwo K^+ K^-$ fit) is used to determine the yield of the $\Bsb
\rightarrow \chi_{c1,c2} K^+ K^-$ components. The $\chicone K^+ K^-$ fit has six free
parameters: the $\Bsb \rightarrow \chicone K^+ K^-$  decay yield,
$N_{\chicone}$,  the $\Bsb \rightarrow \chictwo K^+ K^-$  decay yield
relative to that of the $\chicone$ mode, $f$, the \Bs mass, $m(\Bs)$, the
yield of the partially reconstructed background, $N_{\textrm{part}}$, the
  combinatorial background yield, $N_{\textrm{comb}}$, and the slope of the
  combinatorial background. In addition, $s_f$ is allowed to vary within
  the  Gaussian constraint of $1.00 \pm 0.05$. The  $\chictwo
  K^+ K^-$ fit has the same free parameters apart from
  $m(\Bs)$, which is fixed to its known value
  \cite{PDG2016}. The fit procedure
is validated using both the full simulation and pseudoexperiments which are fits to simulated distributions
generated according to the density functions described above and
using the yields from the fit to the data.  No significant bias is
found and the uncertainties estimated by the fit agree with the
results of the pseudoexperiments.

The results of the fits to the data are shown in Fig.~\ref{fig4} and
the relevant parameters listed in Table \ref{tab:results}. The quality of the fit is
judged to be good from the residuals and by a binned $\chi^2$ test. The value of $N_{\textrm{part}}$ is consistent
with the expectation based on the relevant branching fractions \cite{PDG2016}.
The significance of the $\Bsb \rightarrow \chictwo K^+ K^-$ component, including systematic uncertainties due to the choice
of fit model and evaluated using the fit with $\chictwo \gamma$ mass constraint, is evaluated to be $6.7 \sigma$ using Wilks' theorem
\cite{Wilks:1938dza}. The values of $f$ determined from the two fits
are consistent. That from the $\chictwo K^+ K^-$ fit is more precise
since, as can be seen from Fig.~\ref{fig4}, the width of the $\Bsb \rightarrow \chictwo K^+ K^-$
component is narrower than in the $\Bsb \rightarrow \chicone K^+ K^-$
case. Hence, this value is used in the determination of the ratio of
branching fractions.
\begin{table}[htb!]
\caption{\small Results of the $\chicone K^+ K^-$ and
  $\chictwo K^+ K^-$ fits to the invariant-mass distributions. A Gaussian constraint
is applied to the $s_f$  parameter.}
\begin{center}
\begin{tabular}{l|r@{$\,\pm\,\,$}l|r@{$\,\pm\,\,$}l }
              & \multicolumn{4}{c}{Value} \\
                              \raisebox{1.5ex}[-1.5ex]{Fit parameter}               & \multicolumn{2}{c}{$\chicone$ fit}
                                                              & \multicolumn{2}{c}{$\chictwo$ fit} \\ \hline
$N_{\chicone}$ & $745$ & $30$ & $743$ & $30$\\
$f [\%]$ & $8.3$ & $2.2$  & $10.5$ & $1.9$ \\
$m(\Bs) \, [\mevcc]$ & $5366.83$ & $0.25$ & \multicolumn{2}{c}{---}\\
$N_{\textrm{part}}$ & $390$ & $47$ & $343$ & $46$\\
$N_{\textrm{comb}}$ & $1024$ & $65$ & $1013$ & $62$  \\
$s_f$ & 1.01 & 0.03 & 1.02 & 0.05\\
\end{tabular}
\end{center}
\label{tab:results}
\end{table}

\section{Determination of the {\boldmath $\Bsb \rightarrow \chictwo K^+ K^-$} branching fraction}
\label{sec:bf}
The ratio of branching fractions is calculated as
\begin{equation*}
\frac{\mathcal{B}(\Bsb \rightarrow \chictwo K^+ K^-)}{\mathcal{B}(\Bsb
  \rightarrow \chicone K^+ K^-)} = f \cdot \epsilon_r \cdot \frac{\mathcal{B}(\chicone
  \rightarrow \jpsi \gamma)}{\mathcal{B}(\chictwo  \rightarrow \jpsi
  \gamma)},
\end{equation*}
where $f = (10.5 \pm 1.9) \%$ and
\begin{equation*}
 \frac{\mathcal{B}(\chicone
  \rightarrow \jpsi \gamma)}{\mathcal{B}(\chictwo  \rightarrow \jpsi
  \gamma)}   = 1.77 \pm 0.09,
\end{equation*}
using the values given in Ref. \cite{PDG2016}. The ratio
of reconstruction and selection efficiencies between the two
modes, $\epsilon_r$, is not one due to differences in the photon
kinematics between the two decay modes. It is evaluated using the simulation to be $(92.0
\pm 1.6)\%$ where the uncertainty is statistical. Thus, the ratio of
branching fractions is
\begin{equation*}
\frac{\mathcal{B}(\Bsb \rightarrow \chictwo K^+ K^-)}{\mathcal{B}(\Bsb
  \rightarrow \chicone K^+ K^-)} = (17.1 \pm 3.1)\%,
\end{equation*}
where the uncertainty is statistical.

Since the signal and normalization modes are identical in topology,
systematic uncertainties largely cancel in the ratio of branching
fractions. The assigned systematic uncertainties are listed in Table
\ref{tab:BRsyst}. The limited size of the available simulation samples
leads to a relative uncertainty of $1.8 \%$.  The uncertainty from the
choice of the fit model is evaluated to be $1.5 \%$ using the discrete profiling method described in
Ref. \cite{Dauncey:2014xga}. Propagating the uncertainty on the yield of the
$K^{*}(892)^{0}$ background leads to an additional $0.3\%$ uncertainty.  The effect of
possible differences in the $\Bsb$ kinematics between the data and
simulation is studied by weighting the simulation such that $\pt$
spectra in data and simulation agree for the $\Bsb
\rightarrow \chicone K^+ K^-$ decay mode. Based on this study, a
$0.4\%$ uncertainty is assigned. Summing in quadrature, the total systematic uncertainty amounts to $2.4\%$. No systematic uncertainty is included
for the admixture of $\CP$-odd and $\CP$-even $\Bsb$ eigenstates in the decays,
which is assumed to be the same for both channels \cite{DeBruyn:2012wj}. In the extreme case that
one decay is only from the short-lifetime eigenstate and the other only
from the long-lifetime eigenstate, the ratio would change by
$2.8\%$.

External systematic uncertainties of $3.5\%$ and $3.6\%$ arise from
the knowledge of the radiative $\chicone \rightarrow \jpsi \gamma$ and
$\chictwo \rightarrow \jpsi \gamma$  branching fractions
\cite{PDG2016}. Adding these in quadrature gives an additional uncertainty of $5.0\%$.

\begin{table}[tb]
\centering
\caption{Systematic uncertainties for the measurement of
  the ratio $\mathcal{B}(\Bsb \to \chictwo K^+ K^-)/\mathcal{B}(\Bsb \to \chicone K^+ K^-)$. }
\begin{tabular}{c|c}
Source of systematic uncertainty & Relative uncertainty (\%) \\  \hline
Simulation sample size     & 1.8 \\
Fit model            & 1.5  \\
$\Kstar(892)^0$ background & 0.3 \\
Data/simulation agreement& 0.4 \\ \hline
Sum in quadrature of above              & 2.4 \\\hline
$\mathcal{B}(\chicone \to \jpsi \gamma)$            & 3.5
  \\
$\mathcal{B}(\chictwo \to \jpsi \gamma)$              & 3.6 \\ \hline
Sum in quadrature of external uncertainties             & 5.0 \\
\end{tabular}
\label{tab:BRsyst}
\end{table}

\begin{figure}[tb!]
\begin{center}
\includegraphics[width=0.49\textwidth]{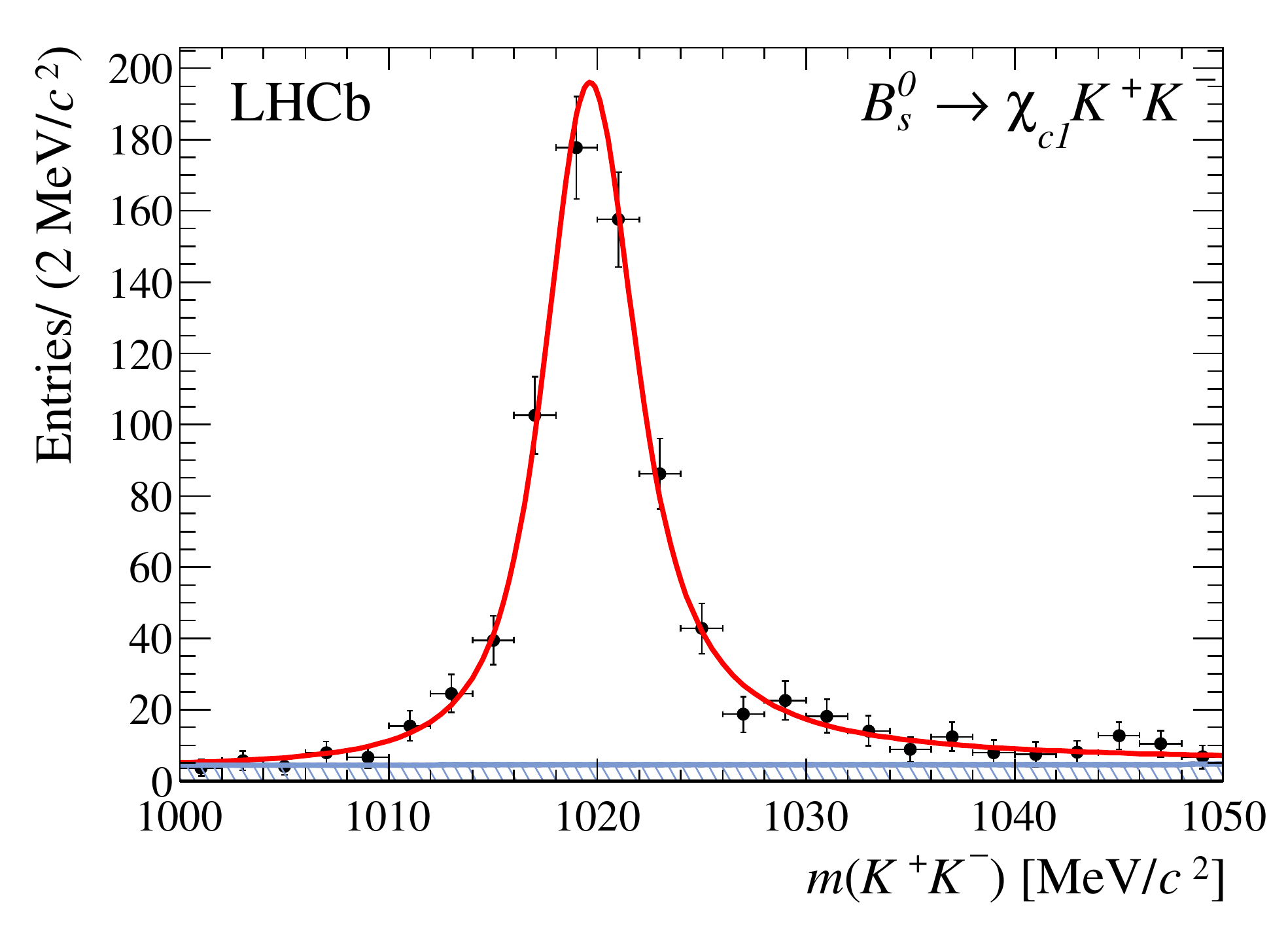}
\includegraphics[width=0.49\textwidth]{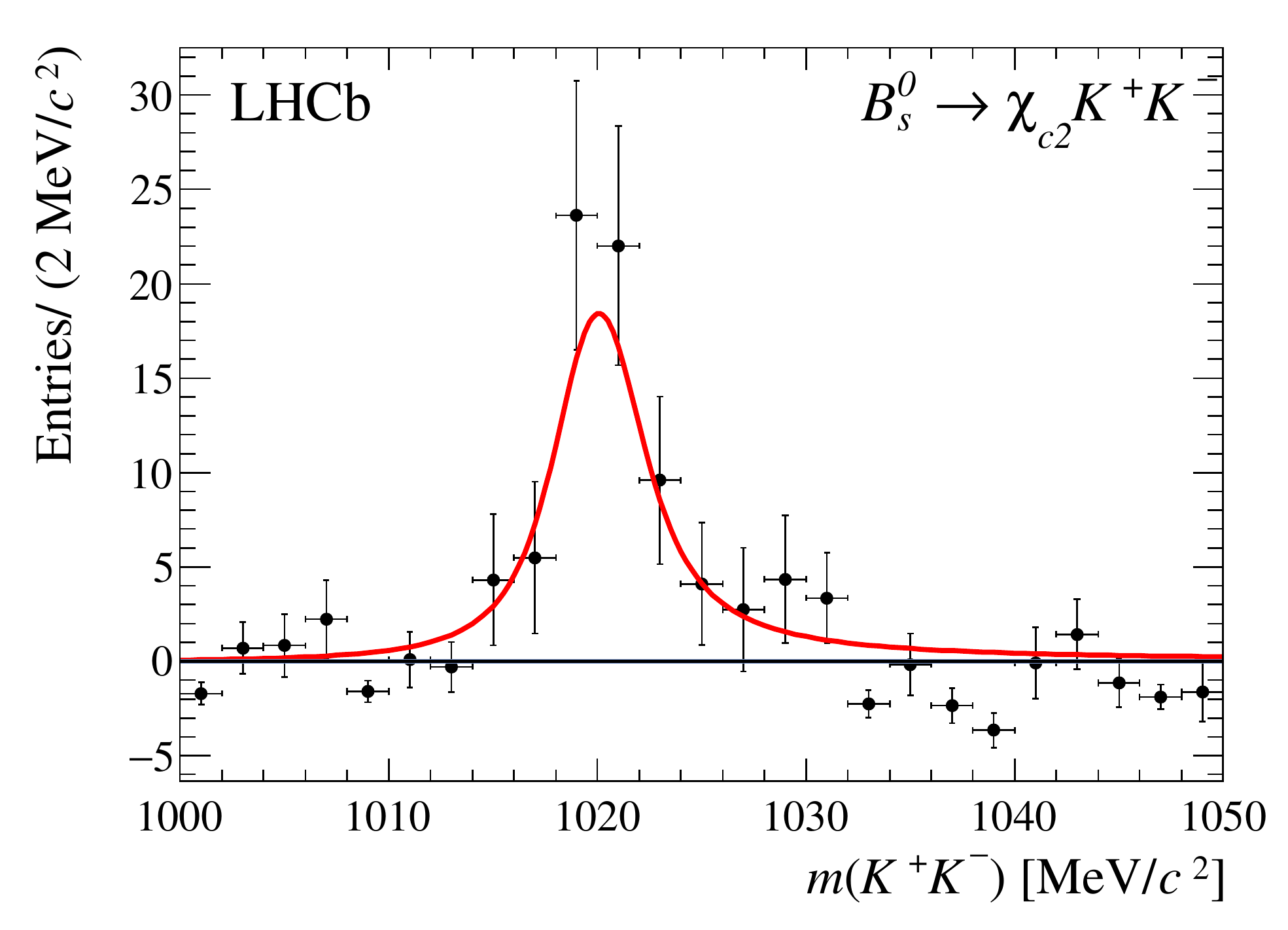}
\caption{\small Invariant-mass distribution of the $K^+K^-$ pair for the (left) $\Bsb
  \rightarrow \chicone K^+ K^-$ decay obtained with the $\chicone$
  mass constraint applied to the $\Bs$ candidate invariant mass and
  (right)  $\Bsb  \rightarrow \chictwo K^+ K^-$ decay obtained with the $\chictwo$
  mass constraint applied to the $\Bs$ candidate invariant mass. The
  (red solid line)  total fitted function is superimposed together with
  (blue hashed area) the S-wave component.}
\label{fig5}
\end{center}
\end{figure}

Both decay modes are expected to be dominated by contributions from an
intermediate $\phi$ resonance that decays to a $K^+K^-$
pair.  Additional S-wave contributions may also be present.  To check
if this is the case, the resonance structure of the $m(K^+K^-)$ invariant-mass
distribution is studied using the \sPlot technique
\cite{Pivk:2004ty}, with weights determined from the $\chicone K^+K^-$
and $\chictwo K^+K^-$ mass fits described in
Sec.~\ref{sec:massfit}. To increase the sensitivity to an S-wave contribution, the $K^+K^-$ mass window $1000$--$1050 \mevcc$ is
considered. The resulting  $K^+K^-$ invariant-mass
distribution is shown in Fig.~\ref{fig5} for the two decay modes.

The observed $K^+K^-$ invariant-mass distribution is modelled with two
components. The first is a relativistic P-wave Breit-Wigner function with Blatt-Weisskopf form factors \cite{Blatt:1952ije}. The
natural width is fixed to the known value of the $\phi$ meson
\cite{PDG2016} and a meson radius
parameter of $3 \hbar c \gev^{-1}$ is used. The detector resolution of
$0.9 \mevcc$ is accounted for by convolving the resonance lineshape with a
Gaussian distribution. The second contribution to the $K^+K^-$
invariant-mass distribution is the S-wave. This is assumed to be
nonresonant in nature and is modelled by a
phase-space function. The fit model has two free parameters:
the $\phi$ mass and the nonresonant S-wave fraction, $f_s$. Applying an unbinned
maximum likelihood fit of this model to the
$\Bsb \rightarrow \chicone K^+ K^-$ sample gives $f_s = (13.9 \pm 2.3)\%$, where the
statistical uncertainty is evaluated using pseudoexperiments. This
value is consistent at the $2 \sigma$-level with that found in the previous
LHCb study \cite{LHCb-PAPER-2013-024} of this mode, $f_s = (3.3 \pm
5.1)\%$. This corresponds to an S-wave fraction of $(9.2 \pm 1.5) \%$ in a
$\pm 15 \mevcc$ window around the $\phi$ mass.

The same procedure is used for the $\Bsb \rightarrow \chictwo K^+ K^-$
sample. In this case the central value of $f_s$ returned by the fit is
zero, that is at the physical boundary. Pseudoexperiments are used to set a limit $f_s < 0.30$ at $90 \%$
confidence level in the $50 \mevcc$ wide $K^+K^-$ mass window. This corresponds to
an S-wave fraction of less than $21 \%$ in a $\pm 15 \mevcc$ window around the $\phi$ mass.

\section{Measurement of the {\boldmath \Bs} mass}
\label{sec:mass}
The fit to the  $\chicone K^+ K^-$ invariant-mass distribution in
Fig.~\ref{fig4} (left) gives ${m(\Bs) = 5366.83 \pm  0.25 \mevcc}$, where the
uncertainty is statistical. The dominant source of systematic
uncertainty on the $\Bs$ mass comes from the knowledge of the momentum
scale for charged-particles. This is found to
be $ 0.26 \mevcc$  by adjusting the momentum scale by the $3 \times 10^{-4}$
uncertainty on the calibration procedure and rerunning the mass fit. A further uncertainty arises from the knowledge of the
amount of material in the spectrometer. This is known to  $10\%$
accuracy \cite{LHCb-DP-2014-002} and results in a $0.02 \mevcc$
uncertainty on the $\Bs$ mass.

The uncertainty from the choice of the fit model is
evaluated to be $0.01 \mevcc$ using the discrete profiling method described in
Ref.~\cite{Dauncey:2014xga}. Finally,
uncertainties of  $0.08 \mevcc$ and $0.02 \mevcc$ arise from the
current knowledge \cite{PDG2016} of the $\chicone$ and $K^+$ masses, respectively.

These uncertainties are summarized in Table~\ref{tab:MassSyst}. Adding them in quadrature results in a systematic
uncertainty of  $0.27 \mevcc$.

\begin{table}[t!]
\caption{\small Systematic uncertainties on the $\Bs$ mass measurement.}
\label{tab:MassSyst}
\begin{center}
\small
\begin{tabular}{l|c}
Source of uncertainty &  Value [$\mevcc$] \\ \hline
Momentum scale                      &         0.26 \\
Material budget             &     0.02   \\
Fit model                        &   0.01 \\
$\chicone$ mass                 &   0.08 \\
$K^+$ mass &  0.02\\
\hline
Sum in quadrature  & 0.27 \\
\end{tabular}
\end{center}
\end{table}

\section{Summary}
\label{sec:summary}
The $\Bsb \rightarrow \chictwo
K^+ K^-$ decay mode is observed for the first time with a significance of $6.7 \sigma$. The
branching fraction of this decay
relative to that of the $\Bsb \rightarrow \chicone K^+ K^-$ mode within a
$\pm 15
\mevcc$ window around the $\phi$ mass is
measured to be
\begin{displaymath}
\frac{\mathcal{B}(\Bsb \rightarrow \chictwo K^+ K^-) }{ \mathcal{B}(\Bsb \rightarrow \chicone K^+ K^- )}
= (17.1 \pm 3.1 \stat \pm  0.4 \syst \pm 0.9 \, (\mathcal{B}))\%.
\end{displaymath}
This ratio agrees with the value measured for the corresponding $\Bz$ decay
by LHCb \cite{LHCb-PAPER-2013-024}
\begin{displaymath}
\frac{\mathcal{B}(\Bz \rightarrow \chictwo K^{*}(892)^0) }{
  \mathcal{B}(\Bz \rightarrow \chicone K^{*}(892)^0)}
= (17.1 \pm 5.0 \stat \pm  1.7 \syst \pm 1.1 \, (\mathcal{B}))\%.
\end{displaymath}
In the $\pm 15 \mevcc$ window around the $\phi$ mass, the nonresonant S-wave fraction for the $\Bsb \rightarrow \chicone
K^+ K^-$ mode
is  measured to be  $(9.2 \pm 1.5)\%$ whilst for the $\Bsb \rightarrow \chictwo K^+ K^-$ mode
it is limited to $<21 \%$ at $90\%$ confidence level.

The $\Bsb \rightarrow \chicone K^+ K^-$ signal is used to measure the
$\Bs$ mass. The result is
\begin{equation*}
m(\Bs) = 5366.83 \pm  0.25\stat \pm 0.27\syst\mevcc\,.
\end{equation*}
This result is in good agreement with and has similar precision to previous LHCb measurements of
the $\Bs$ mass made using the $\Bs
\rightarrow \jpsi \phi$ \cite{LHCb-PAPER-2011-035}  and $\Bs
\rightarrow \jpsi \phi \phi$  \cite{LHCb-PAPER-2015-033} decay modes. The LHCb results are combined, taking the statistical uncertainties
and those related to the fit procedure to be uncorrelated and those
due to the detector material budget and $K^+$ mass to be fully correlated. The uncertainty due to the momentum scale in
Ref. \cite{LHCb-PAPER-2015-033} is also taken to be fully correlated,
whereas in Ref. \cite{LHCb-PAPER-2011-035}  a different procedure was used and so the corresponding uncertainty is considered to be
uncorrelated with the other measurements. The result of this combination is
\begin{equation*}
m(\Bs) = 5366.91 \pm  0.18\stat \pm 0.16\syst\mevcc\,.
\end{equation*}
This value is in good agreement with the value published by the CDF
collaboration, $m(\Bs) = 5366.01 \pm 0.73 \stat \pm 0.33  \syst
\mevcc$ \cite{Acosta:2005mq}, and is the most precise value to date.

\section*{Acknowledgements}
\noindent We express our gratitude to our colleagues in the CERN
accelerator departments for the excellent performance of the LHC. We
thank the technical and administrative staff at the LHCb
institutes. We acknowledge support from CERN and from the national
agencies: CAPES, CNPq, FAPERJ and FINEP (Brazil); MOST and NSFC
(China); CNRS/IN2P3 (France); BMBF, DFG and MPG (Germany); INFN
(Italy); NWO (Netherlands); MNiSW and NCN (Poland); MEN/IFA
(Romania); MinES and FASO (Russia); MinECo (Spain); SNSF and SER
(Switzerland); NASU (Ukraine); STFC (United Kingdom); NSF (USA).  We
acknowledge the computing resources that are provided by CERN, IN2P3
(France), KIT and DESY (Germany), INFN (Italy), SURF (Netherlands),
PIC (Spain), GridPP (United Kingdom), RRCKI and Yandex
LLC (Russia), CSCS (Switzerland), IFIN-HH (Romania), CBPF (Brazil),
PL-GRID (Poland) and OSC (USA). We are indebted to the communities
behind the multiple open-source software packages on which we depend.
Individual groups or members have received support from AvH Foundation
(Germany), EPLANET, Marie Sk\l{}odowska-Curie Actions and ERC
(European Union), ANR, Labex P2IO and OCEVU, and R\'{e}gion
Auvergne-Rh\^{o}ne-Alpes (France), Key Research Program of Frontier
Sciences of CAS, CAS PIFI, and the Thousand Talents Program (China),
RFBR, RSF and Yandex LLC (Russia), GVA, XuntaGal and GENCAT (Spain),
Herchel Smith Fund, the Royal Society, the English-Speaking Union and
the Leverhulme Trust (United Kingdom).

\addcontentsline{toc}{section}{References}
\ifx\mcitethebibliography\mciteundefinedmacro
\PackageError{LHCb.bst}{mciteplus.sty has not been loaded}
{This bibstyle requires the use of the mciteplus package.}\fi
\providecommand{\href}[2]{#2}

\newpage

\centerline{\large\bf LHCb collaboration}
\begin{flushleft}
\small
R.~Aaij$^{27}$,
B.~Adeva$^{41}$,
M.~Adinolfi$^{48}$,
C.A.~Aidala$^{73}$,
Z.~Ajaltouni$^{5}$,
S.~Akar$^{59}$,
P.~Albicocco$^{18}$,
J.~Albrecht$^{10}$,
F.~Alessio$^{42}$,
M.~Alexander$^{53}$
% ...

{\footnotesize \it
% ...
$ ^{5}$Clermont Universit{\'e}, Universit{\'e} Blaise Pascal, CNRS/IN2P3, LPC, Clermont-Ferrand, France\\
% ...
$ ^{10}$Fakult{\"a}t Physik, Technische Universit{\"a}t Dortmund, Dortmund, Germany\\
% ...
$ ^{18}$INFN Laboratori Nazionali di Frascati, Frascati, Italy\\
% ...
$ ^{27}$Nikhef National Institute for Subatomic Physics, Amsterdam, Netherlands\\
% ...
$ ^{41}$Instituto Galego de F{\'\i}sica de Altas Enerx{\'\i}as (IGFAE), Universidade de Santiago de Compostela, Santiago de Compostela, Spain\\
$ ^{42}$European Organization for Nuclear Research (CERN), Geneva, Switzerland\\
% ...
$ ^{48}$H.H. Wills Physics Laboratory, University of Bristol, Bristol, United Kingdom\\
% ...
$ ^{53}$School of Physics and Astronomy, University of Glasgow, Glasgow, United Kingdom\\
% ...
$ ^{59}$University of Cincinnati, Cincinnati, OH, United States\\
% ...
$ ^{61}$Syracuse University, Syracuse, NY, United States\\
% ...
$ ^{73}$University of Michigan, Ann Arbor, United States, associated to $^{61}$\\
% ...
}
\end{flushleft}

\end{document}